# Success rates for linear optical generation of cluster states in coincidence basis

**D B Uskov[1,2*], P M Alsing[3], M L Fanto[3], L Kaplan[2] and A M Smith[3]**

[1]Brescia University, Owensboro, KY 42301
[2]Tulane University, New Orleans, LA 70118
[3]Air Force Research Lab, Information Directorate, 525 Brooks Road, Rome, NY 13440

[*]E-mail: dmitry.uskov@brescia.edu

**Abstract.** We report on theoretical research in photonic cluster-state computing. Finding optimal schemes of generating non-classical photonic states is of critical importance for this field as physically implementable photon-photon entangling operations are currently limited to measurement-assisted stochastic transformations. A critical parameter for assessing the efficiency of such transformations is the success probability of a desired measurement outcome. At present there are several experimental groups which are capable of generating multi-photon cluster states carrying more than eight qubits. Separate photonic qubits or small clusters can be fused into a single cluster state by a probabilistic optical CZ gate conditioned on simultaneous detection of all photons with 1/9 success probability for each gate. This design mechanically follows the original theoretical scheme of cluster state generation proposed more than a decade ago by Raussendorf, Browne and Briegel. The optimality of the destructive CZ gate in application to linear optical cluster state generation has not been analyzed previously. Our results reveal that this method is far from the optimal one. Employing numerical optimization we have identified that maximal success probability of fusing $n$ unentangled dual-rail optical qubits into a linear cluster state is equal to $(1/2)^{n-1}$; $m$-tuple of photonic Bell pair states, commonly generated via spontaneous parametric down-conversion, can be fused into a single cluster with the maximal success probability of $(1/4)^{m-1}$.

## 1. Introduction

Entangled quantum states of light are in great demand in quantum technology today. Photonic quantum information processing, and metrology are all based on exploiting special properties of non-classical multipath entangled states [1]. Due to their high robustness against decoherence, and relatively simple manipulation techniques, photons are often exploited as the primary carriers of quantum information. A generally accepted encoding scheme accepting using photons is the dual rail encoding, in which logical qubit states $|\uparrow\rangle$ and $|\downarrow\rangle$ are encoded in two-mode Fock states $|1,0\rangle$ and $|0,1\rangle$, respectively. In experimental photon implementations, these two modes are commonly associated with horizontal and vertical polarizations. An attractive feature of such an encoding is that single-qubit SU(2) operations can be performed by the standard techniques of linear optics, using practically lossless beam splitters and phase shifters. However, when it comes to entangling photon-encoded qubits, a problem immediately arises: the absence of a photon-photon interaction for coupling the photons.

Optical Kerr nonlinearity can effectively couple photons through their interaction with a dispersive medium. However due to the low photon numbers involved in typical quantum-information processing tasks, such nonlinearity is extremely weak and is of little practical use [2, 3].

Alternatively, an effective photon-photon interaction may be produced using ancilla modes and projective measurements [2-5]. A quantum state generator can then be realized utilizing only linear-optical elements (beam splitters and phase shifters) in combination with photon counters, at the expense of the process becoming probabilistic. The revolutionary discovery by Knill, Laflamme, and Milburn (KLM) [2], which has launched the field of linear optical quantum information processing, was that such a device is capable of transforming an initially separable state into an entangled state. Since the transformation depends on the success of the measurement, the transformation has a probabilistic nature.

The paradigm of quantum computation is based on peculiar laws of quantum mechanics which potentially allow manipulation and processing of information at exponentially faster rates as compared to classical computers. There exist at least two distinct schemes of implementing quantum computation. Historically the first scheme is based on the sequential application of a number of logical gates to elementary carriers of quantum information (qubits).The second scheme, discovered in 2001 by Hans Briegel and Robert Raussendorf [6,7], does not have a classical counterpart: it exploits the purely quantum phenomenon of wave function collapse under a measurement. A computation is performed by inducing non-unitary dynamics in a carefully prepared quantum state of multiple mutually entangled qubits by applying a sequence of measurements according to a desired computational algorithm. Such quantum states are called cluster states or, more generally, quantum graph states.

Since the cluster state paradigm offers better possibilities for error correction this scheme became the leading candidate for the physical realization of quantum information processing. From a physical point of view, photon-based implementations of cluster states, where information is encoded in wave functions of single photons, has important advantages compared to other technologies.

## 2. Optical transformation by postselection in coincidence basis

The quantum measurement-assisted linear optical quantum computer was originally envisioned as a network of linear optical elements (for example in the original KLM scheme), where the controlled sign (C-phase or, equivalently, CZ) gate is constructed as a combination of two nonlinear sign (NS) gates [2]. This approach was effective as a "proof of principle" for linear optical entangling transformations. However, for the technical purpose of building a functional microchip-like device, one does not need to partition the transformation into blocks. Instead, the device may be considered as an "integrated light circuit" [8] which performs one large operation, and one needs to make use of theoretical tools to optimize the fidelity, success, and robustness of the device for a given set of resources available in the form of ancilla photons [4, 9].

First we briefly describe the general scheme of liner-optical transformation.

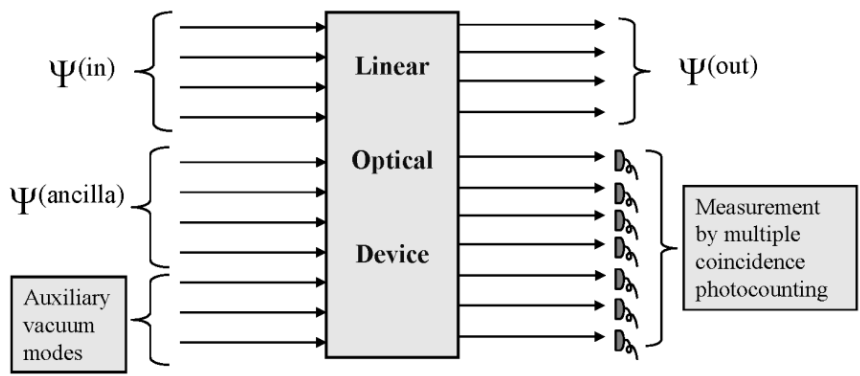

Figure1.A general measurement-assisted transformation using a linear optical interferometer [9]. The "computational" input state $\Psi^{(in)}$ is usually a separable state of two or more dual-rail encoded qubits. The ancilla state is typically assumed to be a separable state. (The actual number of modes is arbitrary for all channels).

The core of the linear optical device is the transformation

$$a_i^\dagger \rightarrow \sum_{j=1}^{N} U_{ij} \tilde{a}_j^\dagger,$$ (1)

of photon creation operators from the input to the output form. Here $N$ is the total number of optical modes, and $U$ is a unitary $N \times N$ matrix, which contains all physical properties of the linear optical device (see Fig. 1). The induced state transformation $\boldsymbol{\Omega}$ is a high-dimensional unitary representation of the matrix $U$. Its action is given by the following algebraic operation on the input state $\left|\Psi^{(total\,input)}\right\rangle = \left|\Psi^{(in)}\right\rangle \left|\Psi^{(ancilla)}\right\rangle$ [10, 11],

$$\left|\Psi^{(total\,output)}\right\rangle = \boldsymbol{\Omega}\left|\Psi^{(total\,input)}\right\rangle = \boldsymbol{\Omega}\prod_{i=1}^{N}\frac{a_i^{\dagger n_i}}{\sqrt{n_i!}}\left|0\right\rangle = \prod_{i=1}^{N}\frac{1}{\sqrt{n_i!}}\left(\sum_j U_{ij}\tilde{a}_j^\dagger\right)^{n_i}\left|0\right\rangle$$ (2)

The map between operators $U \rightarrow \boldsymbol{\Omega}$ is a group homomorphism, i.e. if $U_1 \rightarrow \boldsymbol{\Omega}_1$ and $U_2 \rightarrow \boldsymbol{\Omega}_2$ then $U_1 U_2 \rightarrow \boldsymbol{\Omega}_2 \boldsymbol{\Omega}_1$.

Next, a Von-Neumann measurement in the Fock basis is performed on a subspace of the final state $\left|\Psi^{(total\,output)}\right\rangle$ and only one measurement outcome is accepted as a successful implementation of the transformation. If the measurement involves only the set of ancilla modes, then mathematically this operation is equivalent to projecting the $\left|\Psi^{(total\,output)}\right\rangle$ state onto a predefined Fock state in the ancilla modes $A = \pi r^2 \left|\Psi^{(mesurement)}\right\rangle = \left|k_{N_c+1}, k_{N_c+2}, ...k_N\right\rangle$,

$$\left|\Psi^{(out)}\right\rangle = \left\langle k_{N_c+1}, k_{N_c+2}, ...k_N\left|\boldsymbol{\Omega}\right|\Psi^{(total\,output)}\right\rangle = A\left|\Psi^{(in)}\right\rangle.$$ (3)

Here $A$ is a *contraction* Kraus linear operator [12] acting on the input computational state denoted above as $\left|\Psi^{(in)}\right\rangle$. In the literature transformation (3) is called a measurement-assisted transformation or a Stochastic Local Operations and Classical Communication (the classical communication is important only for schemes utilizing feed-forward technique to increase the success probability).

A special type of measurement-assisted transformation is achieved by merging computational and ancilla modes in the general scheme described above. For these transformations, photons in the computational modes are playing two roles at the same time: i) carriers of quantum information ii) generators of measurement-induced optical nonlinearities. In other words, $\left|\Psi^{(in)}\right\rangle \equiv \left|\Psi^{(total\,input)}\right\rangle$.

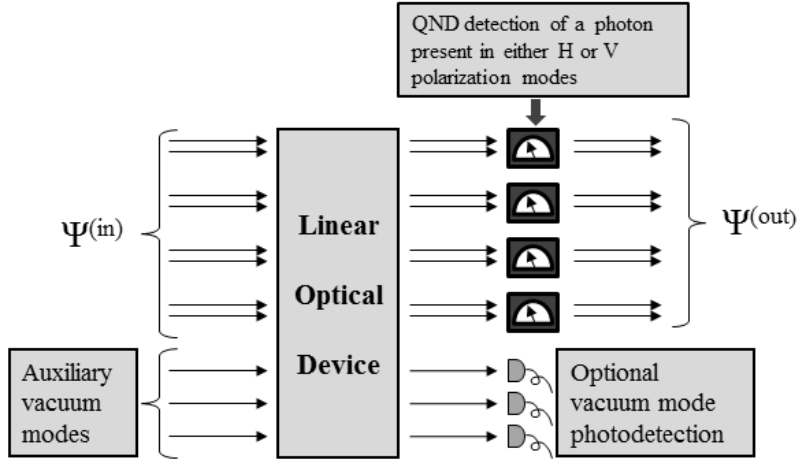

Figure 2. The diagram illustrates the process of state transformation where photo-detection is used to detect the presence of a photon in either one of two modes of a dual-rail pair. Since this type of measurement does not induce a complete collapse of the wave function it will not destroy quantum information encoded in computational dual-rail subspaces including any form of multiqubit entanglement. (The actual number of modes is arbitrary for all channels).

To introduce a proper mathematical description of such schemes we first need to discuss the notion of dual-rail encoding. Contrary to single-rail encoding, when qubit logical states $|0\rangle$ and $|1\rangle$ are encoded in vacuum and one-photon states correspondingly, in dual-rail encoding logical states $|0\rangle$ and $|1\rangle$ are represented by horizontal $|H\rangle$ and vertical $|V\rangle$ polarizations of one spatial mode. From the mathematical point of view horizontal and vertical modes are equivalent to any two orthogonal (spatial)modes since linear optical transformations of photon creation operators of input modes can be equally implemented for polarization rotations and transformations between modes. Therefore dual-rail encoding in general involves two abstract photonic modes and the qubit Hilbert space is simply the single-photon eigenspace of the photon number operator $\hat{N} = a_V^\dagger a_V + a_H^\dagger a_H$. Denoting the states with horizontal and vertical polarization in the $n$-th spatial mode as $|H,V\rangle_n$, the multiqubit space can be formally written as $span(\prod_{n=1}^{N} \otimes |H,V\rangle_n)$. Now instead of a projection on the ancilla state $|\Psi^{(measurement)}\rangle$, as in equation (3), the measurement is represented by the projection operator on the computational subspace $\hat{\mathbf{P}}^{(comp)} = \prod_{n=1}^{N} \left( |H\rangle_n \langle H|_n + |V\rangle_n \langle V|_n \right)$. This is also equivalent to a coincidence photocounting in dual-rail modes. Equation (3) takes the form,

$$\left|\Psi^{(out)}\right\rangle = \hat{\mathbf{P}}^{(comp)} \boldsymbol{\Omega} \left|\Psi^{(in)}\right\rangle = A \left|\Psi^{(in)}\right\rangle . \tag{4}$$

In principle, operation defined by equation (4) implies the application of a quantum non-demolition (QND) device which detects the presence of a photon in two modes without disturbing its quantum state. However, such a requirement can be circumvented in the cluster model of quantum computation when the read-out operation is nothing else but a multiqubit measurement in the basis of qubit product states. Such a simplification imposes certain restrictions on the possibility of concatenating linear optical transformations which we will discuss in detail elsewhere.

Describing the qualitative properties of transformations (4) we would like to clarify the difference between "transformations" and "gates'. The latter is always a unitary operator while the former in general is not (more mathematical details can be found in texts on semi-groups). A special class of transformations (3) or (4) generating matrices $A$ such that $A\,A^\dagger = s\,\hat{I}$, $s \in R$ is called "operational unitary". Here the parameter $s$ is simply the success probability of a unitary gate. In the present paper the focus of our study is on how an arbitrary transformation $A$ acts on specific state, i.e. we are interested only in the action of a transformation on a given input state which is taken to be either a product of single-qubit states or a product of Bell states.

## 3. Fidelity and success probability of optical measurement-assisted transformation of a state

We introduce two important characteristics of state transformation which determine the usefulness of a transformation for generating a desired (or target) state $\left|\psi^{(tar)}\right\rangle$.

The first characteristic of the state transformation quantifies how close the "out-state" $\left|\psi^{(out)}\right\rangle$ is to the target state $\left|\psi^{(tar)}\right\rangle$. It is called the fidelity of the transformation. Mathematically fidelity is defined in terms of the Fubini-Study distance

$$\gamma(\psi^{(out)}, \psi^{(tar)}) = \cos^{-1}\left(\sqrt{\left\langle\psi^{(out)}\middle|\psi^{(tar)}\right\rangle\left\langle\psi^{(tar)}\middle|\psi^{(out)}\right\rangle\middle/\left\langle\psi^{(out)}\middle|\psi^{(out)}\right\rangle\left\langle\psi^{(tar)}\middle|\psi^{(tar)}\right\rangle}\right) \qquad (5)$$

Here $\left|\psi^{(out)}\right\rangle$ is the state given by equation (4). For numerical computations it is expedient to accept the following non-singular parameter as the measure of fidelity: two states have zero distance $\gamma$ if parameter

$$f(U) = \left|\left\langle\psi^{(out)}\middle|\psi^{(tar)}\right\rangle\right|^2\middle/\left\langle\psi^{(out)}\middle|\psi^{(out)}\right\rangle \equiv \cos\left(\gamma\right)^2 = 1 \qquad (6)$$

is equal to one. If the measurement results in a desired outcome the transformation will produce a "collapsed" normalized state $\left|\hat{\psi}^{(out)}\right\rangle = \left|\psi^{(out)}\right\rangle\middle/\sqrt{\left\langle\psi^{(out)}\middle|\psi^{(out)}\right\rangle}$ and the condition of unit fidelity ($f = 1$) guarantees that $\left|\hat{\psi}^{(out)}\right\rangle = e^{i\varphi}\left|\psi^{(tar)}\right\rangle$ (i.e. the target and out-state differ only by a global phase).

In practice the most important characteristic of a measurement-assisted transformation is the value of the success probability of the transformation. While for the gate optimization problem success probability is usually introduced as the Hilbert-Schmidt norm of the operator $A$: $s = Tr(AA^\dagger)/D_c$, where $D_c$ is the dimensionality of the Hilbert space, the success probability of the state transformation can be defined simply as a normalization matrix element

$$s(U) = \left\langle\psi^{(out)}\middle|\psi^{(out)}\right\rangle \equiv \left\langle\psi^{(in)}\middle|A^\dagger A\middle|\psi^{(in)}\right\rangle \equiv \left\langle\psi^{(in)}\middle|\boldsymbol{\Omega}\,\hat{\mathbf{P}}^{(comp)}\boldsymbol{\Omega}\middle|\psi^{(in)}\right\rangle. \qquad (7)$$

The goal of the current study is to find the linear optical matrix $U$ which provides the largest possible success probability $s(U)$ with perfect fidelity ($f(U) = 1$) for generating linear cluster states from single-qubit product states or two-qubit Bell states. Mathematically, both $s(U)$ and $f(U)$ are real-valued functions on the compact SU(N) manifold of unitary operators $U$ and the problem of finding a global maximum of $s(U)$ while keeping perfect fidelity belongs to the category of restricted optimization problems. The numerical implementation of the optimization problem in the present study is similar to technique developed in [11] for gate optimization, where technical details of the optimization code are described. The main feature of the numerical optimization routine which is important for the present paper is that global optimization is pursued by implementing multiple cycles of local optimization with varying starting points and then plotting and analyzing the data for local maxima in the increasing order of success rate.

## 4. $C_4$, $C_6$ and $C_8$ linear cluster state generation from Bell states

To date most experimental research on cluster state generation involves spontaneous parametric down-conversion (SPDC) for producing entangled photon pairs [13-16] (consequently existing schemes are limited to completely stochastic non-heralded generation of cluster states). Cluster state generation is achieved by applying a standard optical CZ gate [17], with the success rate of 1/9. The gate requires two additional vacuum ports i.e. the general scheme in Fig 2 will include two auxiliary vacuum ports. It has been demonstrated by finding direct analytic solution of a set of algebraic equations for transition amplitudes of basis states of two-qubit Hilbert space that the maximal success probability of optical CZ gate is equal to 1/9 [18].

Since the complexity of the problem grows exponentially with the number of qubits involved in the transformation [4], the problem of optimal generation of cluster states cannot be solved analytically even for the problem of generation of $C_4$ state. Therefore we resort to numerical methods developed in [11]. From the point of view of quantum control theory the problem of cluster state generation is the problem of state control rather than control of a quantum transformation acting in a Hilbert space. Therefore full CZ gates may not be the optimal way to generate a cluster state from a specific initial state. This phenomenon has been already confirmed for transformations involving concatenation of several CNOT gates [11] and this idea is being exploited in the present work.

In our studies the input state for generating $C_4$, $C_6$ and $C_8$ linear clusters was taken to be a tensor product of two, three and four Bell states correspondingly $\left|\Psi_4^{(in)}\right\rangle = \left|\Phi^+\right\rangle_{1,2}\left|\Phi^+\right\rangle_{3,4}$, $\left|\Psi_6^{(in)}\right\rangle = \left|\Phi^+\right\rangle_{1,2}\left|\Phi^+\right\rangle_{3,4}\left|\Phi^+\right\rangle_{5,6}$, $\left|\Psi_8^{(in)}\right\rangle = \left|\Phi^+\right\rangle_{1,2}\left|\Phi^+\right\rangle_{3,4}\left|\Phi^+\right\rangle_{5,6}\left|\Phi^+\right\rangle_{7,8}$, where $\left|\Phi^+\right\rangle_{n,m} \equiv \left(\left|H\right\rangle_n\left|H\right\rangle_m + \left|V\right\rangle_n\left|V\right\rangle_m\right)/\sqrt{2}$. We note that Bell State $\left|\Phi^+\right\rangle$ can be morphed into a canonical $C_2$ cluster state $\left|C_2\right\rangle_{n,m} = \left(\left|H\right\rangle_n\left|H\right\rangle_m + \left|H\right\rangle_n\left|V\right\rangle_m + \left|V\right\rangle_n\left|H\right\rangle_m - \left|V\right\rangle_n\left|V\right\rangle_m\right)/2$ by a deterministic local unitary transformation acting on polarization modes $a_{H,m}^\dagger \rightarrow \left(\tilde{a}_{H,m}^\dagger + \tilde{a}_{V,m}^\dagger\right)/\sqrt{2}$, $a_{V,m}^\dagger \rightarrow \left(\tilde{a}_{H,m}^\dagger - \tilde{a}_{V,m}^\dagger\right)/\sqrt{2}$. The target states are taken to be a canonical cluster states generated by applying an abstract two-qubit entangling CZ gate between neighboring qubits prepared in so-called "plus" states $\left|\psi^+\right\rangle = \left(\left|H\right\rangle + \left|V\right\rangle\right)/\sqrt{2}$.

Our numerical results show excellent convergence to a global maximum (see Figure 3 below). In principle, linear optical transformations may be extended to a broader class of non-unitary matrices $U$. The subsequent implementation of such a matrix in the form of a linear optical device requires dilation of a non-unitary matrix to a unitary matrix by adding extra modes, called vacuum modes (i.e. modes carrying zero input photons as shown in see Fig. 2). Our search in the space of non-unitary matrices shows that solutions with success probability larger than 0.16 are automatically unitary.

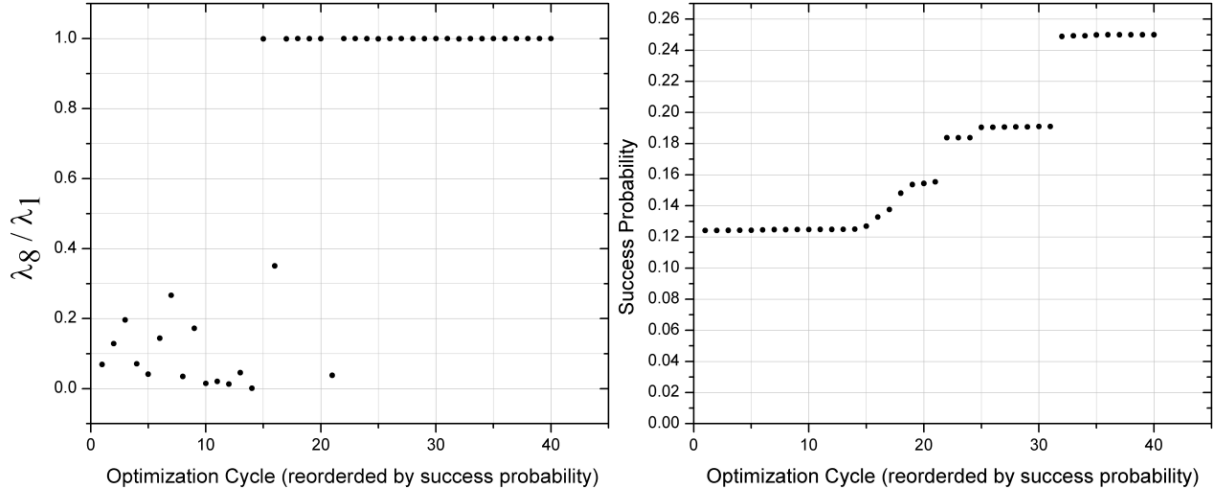

Figure 3a (left panel). The ratio of the smallest singular value of a transformation matrix to the largest singular value. If the ratio is equal to 1 then all singular values are equal to one and the matrix is unitary. These results are obtained for generation of $C_4$ cluster state from a pair of two Bell states.

Figure 3b (right panel). Success probability for generation of linear $C_4$ cluster state from a pair of Bell states: a sample of 40 optimization cycles reordered by increasing success probability (overall we accumulated statistics for more than 2000 cycles confirming that s=1/4 is the global maximum).

The result in Figure 3b immediately demonstrates that the standard scheme of cluster state generation using a destructive CZ gate to fuse two photonic Bell states is not optimal. The success probability can be improved by a factor of 9/4 by modifying the linear optical part of the experimental setup. Our next results for $C_6$ and $C_8$ are shown in Figure 4.

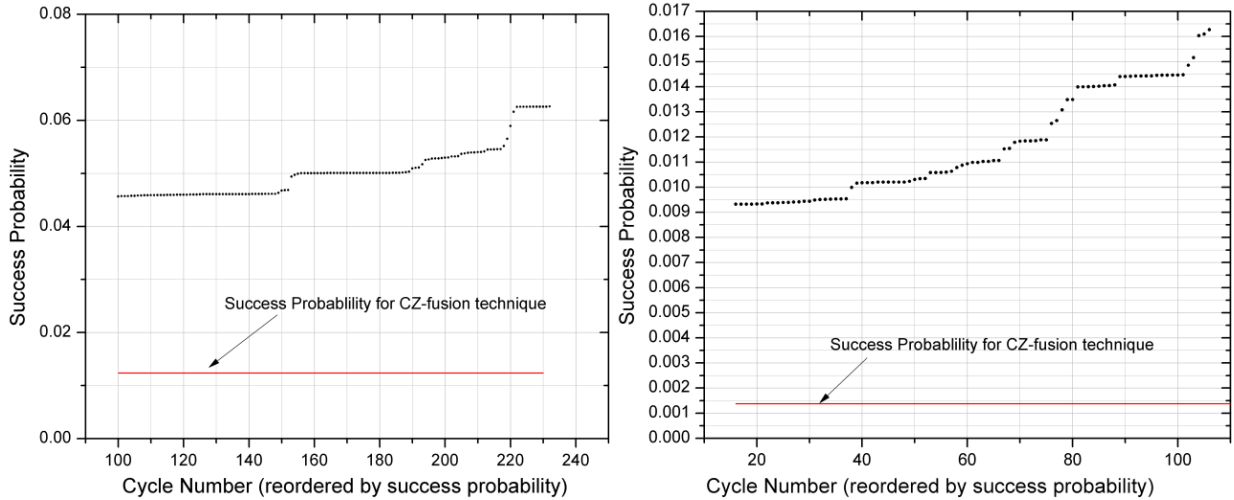

Figure 4a (left panel). Optimization of success probability of $C_6$ cluster state generation from three Bell states. The y-axis shows the value of success probability. The maximal success probability s=$(1/4)^2$.

Figure 4b (right panel). Results of the optimization of the $C_8$ cluster state generation from three Bell state pairs of photonic qubits. The y-axis shows the value of success probability. The maximal success probability is numerically close to $s = (1/4)^3$. Numerical values of fidelity and success probability for best three points are s= {0.016032, 0.016096, 0.016272} and f={0.999997, 0.999998, 0.999994} correspondingly.

We observe that the maximal success probability for the $C_6$ cluster state is numerically very close to $0.0625 = 1/4^2$. Since the numerical complexity of multiphoton optimization problem grows factorially with the number of photons we were able to find only a few local maxima for the $C_8$ cluster state generation problem. However the general trend of numerical results for the set of $C_4$, $C_6$ and $C_8$ states indicates that the maximal success probability for generation of a $C_n$ linear cluster state (here n indicates the number of qubits in a cluster) from $n/2$ photonic Bell states depends on the number of additional fusion links $m = n/2 - 1$ as $1/4^m$. For the $C_4$ cluster state this number is 1, for the $C_6$ cluster state m=2 and for the $C_8$ cluster state m=3. Due to the increasing numerical complexity of global optimization and we were not able to verify this result for cluster states larger than $C_8$. However, our results demonstrate that the computational advantage of the optimal scheme grows exponentially fast with the size of the cluster state. For the $C_4$ cluster state generation we obtain a factor of 9/4 improvement compared with the traditional scheme; for the $C_6$ cluster state this factor is $(9/4)^2 \approx 5$; and for the $C_8$ cluster state the improvement factor is $(9/4)^3 \approx 11$. Based upon these results we expect that for higher-dimensional states the advantage of the optimal scheme will continue to grow as a power of $9/4$.

### 5. C4, C6 and C8 cluster state generation from product states

In this section we describe the numerical results of fidelity constrained optimization of the success probability by presenting the data on optimization of success probability for cluster state generation from product states. In contrast with the previous section the initial state does not contain any "pre-loaded" entanglement as in Bell states and one expects that success probability of generating a cluster state, where entanglement permeates the whole cluster, will be reduced. Surprisingly this common-sense reasoning turns out to be incorrect. Our results can be concisely formulated as follows: the success probability of generating a $C_n$ cluster state from n-photon product states is only a factor of 2 smaller than maximal success probability of generating a $C_n$ cluster state from $n/2$-tuple of Bell states.

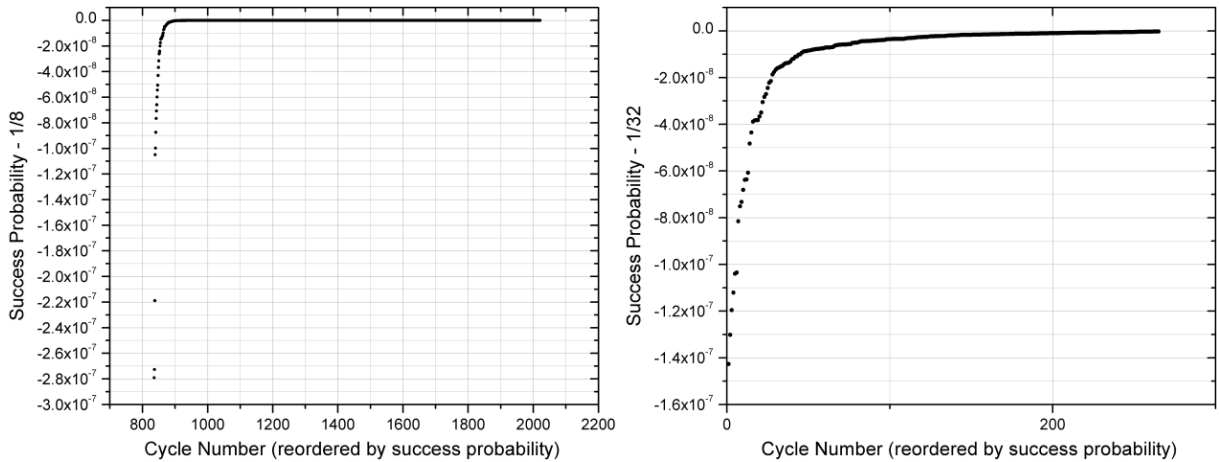

Figure 5a (left panel). $C_4$ cluster state generated from a product state. The y-axis shows the value of success probability minus its maximal value of 1/8.

Figure 5b (right panel). $C_6$ cluster state generated from a product state. The y-axis shows the value of success probability minus its maximal value of 1/32.

Unfortunately the structure of optimal solutions encoded in the linear optical matrix $U$ is too complicated to allow simple analysis of the underlying mechanisms of this phenomenon for general solutions.

However power-law dependence of the success rate of optimal fusion transformations strongly suggests that in the process of fusing a Bell state to a $C_n$ cluster the entanglement of the Bell state does not help to increase the success probability of the operation. In other words, we expect that sequential fusion of two unentangled single-photon states into a $C_n$ cluster state, resulting in $C_m$ cluster state with $m = n + 2$, can be implemented with the same efficiency as fusion of a Bell state to $C_n$ cluster state.

| | $C_2$ | $C_3$ | $C_4$ | $C_5$ | $C_6$ | $C_7$ | $C_8$ |
|---|---|---|---|---|---|---|---|
| From Bell Pairs (optimal) | 1 | n/a | 1/4 | n/a | 1/16 | n/a | 1/64 |
| From n-qubit Product state (optimal) | 1/2 | 1/4 | 1/8 | 1/16 | 1/32 | 1/64 | 1/128 |
| Fusion by CZ gate from product state | 1/9 | 1/81 | 1/729 | 1/6561 | 1/59049 | 1/531441 | 1/4782969 |
| Fusion by CZ gate from Bell pairs | n/a | n/a | 1/9 | n/a | 1/81 | n/a | 1/729 |

Table 1. Results of success probability of cluster state generation (combined by results for $C_3$ and $C_5$ clusters).

## 6. Conclusions

We performed a numerical analysis of the problem of photonic cluster-state generation in application to quantum optics. Our method performs a search for the most efficient scheme of cluster state generation from either a combination of untangled photons or a set of pairs of entangled photons.

The optimization tasks we performed are of critical importance for photonic quantum computation since the only photon-photon entangling operations currently implementable with high repetition rate and fidelity are measurement-assisted stochastic quantum transformations. Our results demonstrate that standard methods of cluster state generation using standard probabilistic linear optical CZ (C-phase) gate is far from optimal. Performing numerical optimization we established that there exists a scheme of cluster state generation which boosts the success rate of generation by more than an order of magnitude even for a small eight-qubit cluster state. The advantage of this scheme in comparison with traditional schemes grows exponentially fast with the size of a cluster.

Finding the simplest possible realization of the scheme with the fewest number of optical elements requires further analytical and numerical analysis.

## Acknowledgements

DBU acknowledges support from AFRL Information Directorate under grant FA 8750-11-2-0218 and from NSF under grant PHY 1005709. Any opinions, findings, and conclusions or recommendations expressed in this material are those of the authors and do not necessarily reflect the views of AFRL.